\documentclass[preprint,aps]{revtex4}
\usepackage{amssymb,epsf}

\begin{document}

\title{Entropic Corrections to Coulomb's Law}
\author{S. H. Hendi$^{1,2}$\footnote{E-mail: hendi@mail.yu.ac.ir} and A. Sheykhi$^{2,3} $\footnote{E-mail: sheykhi@mail.uk.ac.ir}}
\affiliation{$^1$ Physics Department, College of Sciences, Yasouj University, Yasouj 75914, Iran\\
$^2$ Research Institute for Astronomy and Astrophysics of Maragha
(RIAAM), Maragha, Iran\\ $^3$ Department of Physics, Shahid
Bahonar University, Kerman 76175, Iran}

\begin{abstract}
Two well-known quantum corrections to the area law have been
introduced in the literatures, namely, logarithmic and power-law
corrections. Logarithmic corrections, arises from loop quantum
gravity due to thermal equilibrium fluctuations and quantum
fluctuations, while, power-law correction appears in dealing with
the entanglement of quantum fields in and out the horizon.
Inspired by Verlinde's argument on the entropic force, and
assuming the quantum corrected relation for the entropy, we
propose the entropic origin for the Coulomb's law in this note.
Also we investigate the Uehling potential as a radiative
correction to Coulomb potential in $1$-loop order and show that
for some value of distance the entropic corrections of the
Coulomb's law is compatible with the vacuum-polarization
correction in QED. So, we derive modified Coulomb's law as well as
the entropy corrected Poisson's equation which governing the
evolution of the scalar potential $\phi$. Our study further
supports the unification of gravity and electromagnetic
interactions based on the holographic principle.
\end{abstract}

\maketitle

\section{Introduction}

The profound connection between gravity and thermodynamics has a
long history since the discovery of black holes thermodynamics in
1970's \cite{HB,B,D}. It was discovered that black holes can emit
Hawking radiation with a temperature proportional to its surface
gravity at the black hole horizon and black hole has an entropy
proportional to its horizon area \cite{B}. The Hawking temperature
and horizon entropy together with the black hole mass obey the
first law of black hole thermodynamics \cite{D}. The studies on
the connection between gravity and thermodynamics has been
continued until in 1995 Jacobson showed that the Einstein field
equation is just an equation of state for spacetime and in
particular it can be derived from the the first law of
thermodynamics together with relation between the horizon area and
entropy \cite{Jac}. Following Jacobson, however, several recent
investigations have shown that there is indeed a deeper connection
between gravitational dynamics and horizon thermodynamics. It has
been shown that the gravitational field equations in a wide
variety of theories, when evaluated on a horizon, reduce to the
first law of thermodynamics and vice versa. This result, first
pointed out in \cite{Pad1}, has now been demonstrated in various
theory including f(R) gravity \cite{Elin}, cosmological setups
\cite{Cai2,Cai3,CaiKim,Wang,Cai33,Shey0}, and in braneworld
scenarios \cite{Shey1,Shey2}. For a recent review on the
thermodynamical aspects of gravity and complete list of references
see \cite{Padrev}. Although Jacobson's derivation is logically
clear and theoretically sound, the statistical mechanical origin
of the thermodynamic nature of gravity remains obscure.

\begin{figure}[tbp]
\epsfxsize=6cm \centerline{\epsffile{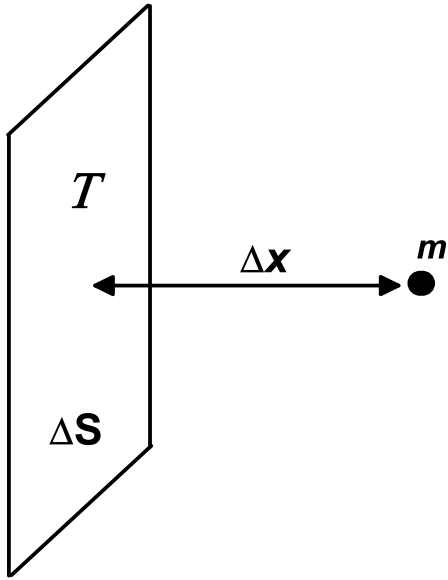}} \caption{Test
particle with mass $m$ approaches a holographic screen with the
temperature $T$ and the entropy change $\triangle S$.} \label{Rec}
\end{figure}

A constructive new idea on the relation between gravity and
thermodynamics was recently proposed by Verlinde \cite{Ver} who
claimed that gravity is not a fundamental interaction and can be
interpreted as an entropic force arising from the change of
information when a material body moves away from the holographic
screen. Verlinde postulated that when a test particle approaches a
holographic screen from a distance $\triangle x$, the magnitude of
the entropic force on this body has the form
\begin{equation}  \label{F}
F\triangle x=T \triangle S,
\end{equation}
where $T$ and $\triangle S$ are the temperature and the entropy
change on the screen, respectively (see Fig. \ref{Rec}).

Focusing on the physical explanation of interesting proposal of
Verlinde, it has been shown that his idea is problematic
\cite{Hossenfelder2010,Gao2011}. In other word, although
Verlinde's derivation is right, mathematically, it does not prove
that gravity is an entropic force, physically. We should note that
it has been presented a general objection to viewing gravity as an
entropic force \cite{Gao2011} and it has been proved that
Verlinde's idea is supported by a mathematical argument based on a
discrete group theory \cite{Winkelnkemper}. In addition,
considering a modified entropic force with the covariant entropy
bound, one may obtain the Newtonian force law \cite{Myung2011}.
Also, following the controversial hypothesis in Ref.
\cite{Chaichian2011}, it has been shown that gravity is an
entropic force.

Verlinde's derivation of laws of gravitation opens a new window to
understand gravity from the first principles. The entropic
interpretation of gravity has been used to extract Friedmann
equations at the apparent horizon of the
Friedmann-Robertson-Walker universe \cite{Cai4},
modified Friedmann equations \cite{Sheykhi}, modified Newton's law \cite%
{Modesto}, the Newtonian gravity in loop quantum gravity
\cite{smolin}, the holographic dark energy \cite{Mli},
thermodynamics of black holes \cite{Tian} and the extension to
Coulomb force \cite{Twang}. Other studies on the entropic force
have been carried out in \cite{Other}.

In addition, the derivation of Newton's law of gravity, in
Verlinde's approach, depends on the entropy-area relationship
$S=A/4\ell _{p}^{2}$ of black holes in Einstein's gravity, where
$A=4\pi R^{2}$ represents the area of the horizon and $\ell
_{p}^{2}=G\hbar /c^{3}$ is the Planck length. However, this
definition can be modified from the inclusion of quantum effects.
Two well-known quantum corrections to the area law have been
introduced in the literatures, namely, logarithmic and power-law
corrections. Logarithmic corrections, arises from loop quantum
gravity due to thermal equilibrium fluctuations and quantum
fluctuations \cite{Meis,Zhang},
\begin{equation}
S=\frac{A}{4\ell _{p}^{2}}-\beta \ln {\frac{A}{4\ell
_{p}^{2}}}+\gamma \frac{\ell _{p}^{2}}{A}+\mathrm{const},
\label{S1}
\end{equation}
where $\beta $ and $\gamma $ are dimensionless constants of order
unity. The exact values of these constants are not yet determined
and still an open issue in quantum gravity.

Power-law correction appears in dealing with the entanglement of
quantum fields in and out the horizon. The entanglement entropy of
the ground state obeys the Bekenstein- Hawking area law. However,
a correction term proportional to a fractional power of area
results when the field is in a superposition of ground and excited
states \cite{Sau}. In other words, the excited state contributes
to the power-law correction, and more excitations produce more
deviation from the area law \cite{sau1}. The power-law corrected
entropy is written as \cite{Sau,pavon1}
\begin{equation}
S=\frac{A}{4\ell _{p}^{2}}\left[ 1-K_{\alpha }A^{1-\alpha
/2}\right] \label{plec}
\end{equation}%
where $\alpha $ is a dimensionless constant whose value ranges as $%
2<\alpha<4 $ \cite{Sau}, and
\begin{equation}
K_{\alpha }=\frac{\alpha (4\pi )^{\alpha /2-1}}{(4-\alpha
)r_{c}^{2-\alpha }} \label{kalpha}
\end{equation}%
where $r_{c}$ is the crossover scale. The second term in Eq.
(\ref{plec}) can be regarded as a power-law correction to the area
law, resulting from entanglement, when the wave-function of the
field is chosen to be a superposition of ground state and exited
state \cite{Sau}. Taking the corrected entropy-area relation into
account, the corrections to the Newton's law of gravitation as
well as the modified Friedman equations were derived
\cite{Sheykhi}.

In this paper, we would like to extend the study to the
electromagnetic interaction. We will derive the general quantum
corrections to the Coulomb's law, Poisson's equation and the
general form of the modified Newton-Coulomb's law by assuming the
entropic origin for the electromagnetic interaction.

\section{Entropic corrections to Coulomb's law}

\begin{figure}[tbp]
\epsfxsize=8cm \centerline{\epsffile{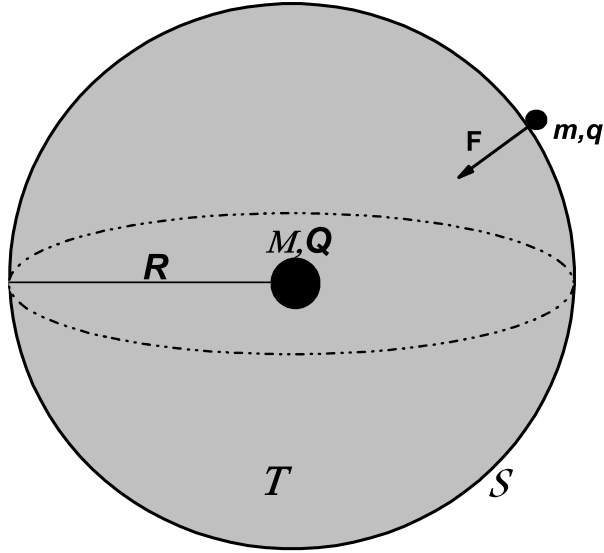}} \caption{A test
charged particle with mass $m$ and charge $q$ and the other
considered as the source with respective charge $Q$ and mass $M$
surrounded by the spherically symmetric screen $\mathcal{S}$.}
\label{Spherical}
\end{figure}

In order to derive the corrections to the Coulomb's law of
electromagnetic, we consider the modified entropy-area
relationship in the following form
\begin{equation}
S=\frac{A}{4\ell _{p}^{2}}+{s}(A),  \label{S2}
\end{equation}%
where $s(A)$ represents the general quantum correction terms in
the entropy expression. We assume there are two charged particles,
one a test charged particle with mass $m$ and charge $q$ and the
other considered as the source with respective charge $Q$ and mass
$M$ located at the center (see Fig. \ref{Spherical} for more
details). Centered around the source mass $M$ with charge $Q$, is
a spherically symmetric surface $\mathcal{S}$ which will be
defined with certain properties that
will be specified explicitly later. To derive the entropic law, the surface $%
\mathcal{S}$ is between the test mass and the source mass, but the
test mass is assumed to be very close to the surface as compared
to its reduced Compton wavelength $\lambda _{m}=\frac{\hbar
}{mc}$. When a test mass $m$ is a distance $\triangle x=\eta
\lambda _{m}$ away from the surface $\mathcal{S} $, the entropy of
the surface changes by one fundamental unit $\triangle S$ fixed by
the discrete spectrum of the area of the surface via the relation
\begin{equation}
\triangle S=\frac{\partial S}{\partial A}\triangle A=\left(
\frac{1}{4\ell _{p}^{2}}+\frac{\partial {s}(A)}{\partial A}\right)
\triangle A.  \label{S3}
\end{equation}
We find out that in order to interpret the entropic origin for the
electromagnetic force, we should leave away the relativistic rest
mass energy $E=Mc^{2}$, and instead, in a similar manner, we
propose the relativistic rest electromagnetic energy of the source
$Q$ as
\begin{equation}
E=\Gamma Q c^{2},  \label{Ec}
\end{equation}
where $\Gamma =\chi q/m$, and $\chi $ is a constant with known
dimension ($[\chi ]=\frac{[k]}{[G]}$, where $k$ and $G$ are
Coulomb and Newtonian constants, respectively). Although the
physical interpretation of assumption (\ref{Ec}) is not clear well
yet for us, however as we will see it leads to the reasonable
results. It is notable to mention that the charge/mass ratio
($q/m$) is a physical quantity that is widely used in the
electrodynamics of charged particles. When a charged particle
follows a circular which is caused by the magnetic field, the
magnetic force is acting like a centripetal force. It is easy to
find that charge/mass ratio ($q/m=V/Br$) is a constant in which we
equal it to $\Gamma/\chi $.

Considering the relativistic rest mass energy relation with the
motivation of analogy between mass in gravity and charge in
electromagnetic interactions, one may consider
$E_{EM}=\mathcal{M}_{EM}c^{2}$, in which $E_{EM}$ is the
electromagnetic energy and $\mathcal{M}_{EM}=\Gamma Q$ is its
corresponding mass which we call it as the electromagnetic mass.
It is notable that there are other concepts of mass in special
relativity, such as longitudinal mass and transverse mass.

We should mention that we are working in the geometrized unit of
charge, in which the Coulomb's law takes almost the same form as
the Newton's law except for the difference in signature. On the
surface $\mathcal{S}$, there live a set of \textquotedblleft
bytes" of information that scale proportional to the area of the
surface so that
\begin{equation}
A=\xi N,  \label{AQN}
\end{equation}
where $N$ represents the number of bytes and $\xi $ is a
fundamental constant which should be determined later. Assuming
the temperature on the surface is $T$, and then according to the
equipartition law of energy \cite{Pad3}, the total energy on the
surface is
\begin{equation}
E=\frac{1}{2}Nk_{B}T.  \label{E}
\end{equation}
Finally, we assume that the electric force on the charge particle
$q$ follows from the generic form of the entropic force governed
by the thermodynamic equation of state
\begin{equation}
F=T\frac{\triangle S}{\triangle x},  \label{F2}
\end{equation}
where $\triangle S$ is one fundamental unit of entropy when
$|\triangle x|=\eta \lambda _{m}$, and the entropy gradient points
radially from the outside of the surface to inside. Note that $N$
is the number of bytes and thus $\triangle N=1$; hence from
(\ref{AQN}) we find $\triangle A=\xi $. Now, we are in a position
to derive the entropy-corrected Coulomb's law. Combining Eqs.
(\ref{S3})- (\ref{F2}), we reach
\begin{eqnarray}
F &=&\frac{2\Gamma Qc^{2}}{Nk_{B}}\frac{\Delta A}{\Delta x}\left(
\frac{
\partial S}{\partial A}\right)  \nonumber \\
&=&\frac{2\Gamma Q\xi mc^{3}}{Nk_{B}\eta \hbar }\left(
\frac{\partial S}{
\partial A}\right)  \nonumber \\
&=&\frac{Qq}{R^{2}}\left( \frac{\chi \xi ^{2}c^{3}}{8\pi k_{B}\eta
\hbar \ell _{p}^{2}}\right) \left[ 1+4\ell _{p}^{2}\frac{\partial
{s}}{\partial A} \right] _{A=4\pi R^{2}},  \label{F3}
\end{eqnarray}
This is nothing but the Coulomb's law of electromagnetic to the
first order provided we define $\xi ^{2}=8\pi k_{B}\eta \ell
_{p}^{4}$ and $\chi =1/(4\pi \varepsilon _{0}G)=\hbar /(4\pi
\varepsilon _{0}\ell _{p}^{2}c^{3})$ . Thus we write the general
quantum corrected Coulomb's law as
\begin{equation}
F_{\mathrm{em}}= \frac{1}{4 \pi \varepsilon
_{0}}\frac{Qq}{R^{2}}\left[ 1+4\ell _{p}^{2}\frac{\partial {s}}{
\partial A}\right] _{A=4\pi R^{2}}.  \label{F4}
\end{equation}
In order to specify the correction terms explicitly, we use the
two well-known kinds of entropy corrections. It is easy to show
that
\begin{equation}
F_{\mathrm{em1}}=\frac{1}{4 \pi \varepsilon _{0}}
\frac{Qq}{R^{2}}\left[ 1-\frac{\beta }{\pi }\frac{\ell
_{p}^{2}}{R^{2}}-\frac{\gamma }{4\pi ^{2}}\frac{\ell
_{p}^{4}}{R^{4}}\right] ,  \label{F5}
\end{equation}
\begin{equation}
F_{\mathrm{em2}}=\frac{1}{4 \pi \varepsilon _{0}}
\frac{Qq}{R^{2}}\left[ 1-\frac{\alpha }{2}\left(
\frac{r_{c}}{R}\right) ^{\alpha -2}\right] ,  \label{F55}
\end{equation}
where $F_{\mathrm{em1}}$ and $F_{\mathrm{em2}}$ are, respectively,
the logarithmic and power-law corrected Coulomb's law. Thus, with
the corrections in the entropy expression, we see that the
Coulomb's law will be modified accordingly. Since the correction
terms in Eqs. (\ref{F5}) and (\ref{F55}) can be comparable to the
first term only when $R$ is very small (i.e. $R<<l_{p}$ and
$R<<r_{c}$ for Eqs. (\ref{F5}) and (\ref{F55}), respectively), the
corrections make sense only at the very small distances (note that
$\alpha >2$). For large distances (i.e. $R>>l_{p}$ for (\ref{F5})
and $R>>r_{c}$ for (\ref{F55})), the entropy-corrected Coulomb's
law reduces to the usual Coulomb's law of electromagnetic.

\section{Uehling Correction to Coulomb's law}

In order to compare the entropic correction with QED correction of
Coulomb's law, we introduce the the so called Uehling potential
\cite{Uehling} as a radiative correction to Coulomb potential in
$1$-loop order (the vacuum-polarization correction for an electron
in a nuclear Coulomb field). Using the Born approximation, the
relation between the scattering amplitude $M$ and the potential is
given by
\begin{equation}
<p^{\prime }\left\vert iM\right\vert p>=-i2\pi V(\mathbf{q})\delta
(E_{p^{\prime }}-E_{p}),  \label{Ampl}
\end{equation}
where $p$ ($p^{\prime }$) and $E_{p}$ ($E_{p^{\prime }}$) are the
momenta and energy of the incoming (outgoing) particles,
respectively, and $\mathbf{q}=\mathbf{p}^{\prime}-\mathbf{p}$. For
ordinary QED, the amplitude of a particle-antiparticle scattering
is given by \cite{Peskin}
\begin{equation}
iM\sim -\frac{ie^{2}}{\left\vert \mathbf{p}^{\prime
}-\mathbf{p}\right\vert ^{2}}.  \label{iM}
\end{equation}
Comparing (\ref{iM}) with (\ref{Ampl}), one can show that the
attractive classical Coulomb potential $V(\mathbf{q})$ is given by
\begin{equation}
V(\mathbf{q})=-\frac{e^{2}}{\left\vert \mathbf{q}\right\vert
^{2}}, \label{Vq}
\end{equation}
where $\left\vert \mathbf{q}\right\vert =\left\vert
\mathbf{p}-\mathbf{p}^{\prime }\right\vert $. Using a Fourier
transformation into the coordinate space, one can find
\begin{equation}
V(\mathbf{x})=\int \frac{d^{3}q}{(2\pi
)^{3}}V(\mathbf{q})e^{i\mathbf{q}. \mathbf{x}}=-\frac{\alpha
^{\prime }}{\mathbf{R}},  \label{Vx1}
\end{equation}
where $\mathbf{R}=|\mathbf{x}|$ and $\alpha ^{\prime }$ is the
fine structure constant. Furthermore, to include the quantum
correction into the result, the modified Coulomb potential can be
calculated from
\begin{equation}
V(\mathbf{x})=-e^{2}\int \frac{d^{3}q}{(2\pi
)^{3}}\frac{e^{i\mathbf{q}. \mathbf{x}}}{\mathbf{q}^{2}[1-\Pi
(\mathbf{q}^{2})]},  \label{Vx2}
\end{equation}
where $\Pi (\mathbf{q})$ in the ordinary QED is defined by the
vacuum polarization tensor
\begin{equation}
\Pi ^{\mu \nu }(q)=(q^{2}g^{\mu \nu }-q^{\mu }q^{\nu })\Pi
(q^{2}), \label{Pimn}
\end{equation}
and is given by
\begin{equation}
\Pi (q^{2})=-\frac{2\alpha ^{\prime }}{\pi
}\int\limits_{0}^{1}x(1-x)\log \left(
\frac{m^{2}}{m^{2}-x(1-x)q^{2}}\right) dx.  \label{Pi}
\end{equation}
Choosing $q_{0}=0$ and inserting this relation into (\ref{Vx2}),
after some straightforward calculation \cite{Peskin}, one can
obtain the so called Uehling potential
\begin{equation}
V(R)=-\frac{\alpha ^{\prime }}{R}\left( 1+\frac{\alpha ^{\prime
}}{4\sqrt{\pi }}\frac{e^{-2mR}}{(mR)^{3/2}}+...\right) ,
\label{Uehling}
\end{equation}
and after differentiation we can obtain the corresponding Uehling
force
\begin{equation}
F_{Ueh}=\frac{\alpha ^{\prime }}{R^{2}}\left( 1-\frac{\alpha
^{\prime }}{8 \sqrt{\pi
}}\frac{e^{-2mR}}{(mR)^{3/2}}(4mR+5)+...\right) . \label{UehlingF}
\end{equation}
\begin{figure}[tbp]
\epsfxsize=12cm \centerline{\epsffile{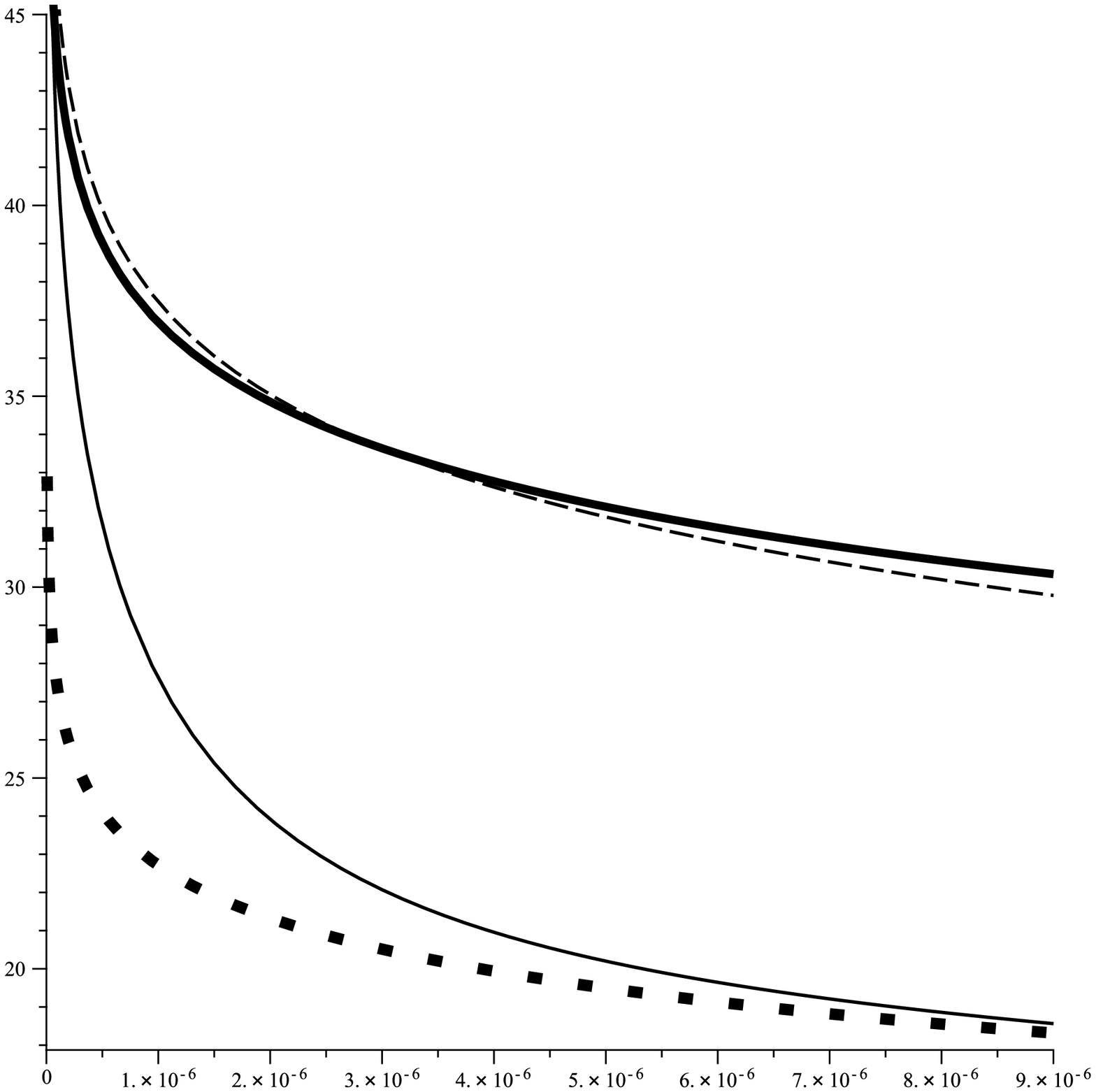}} \caption{$\ln
F_{em1}$ (solid line), $\ln F_{em2}$ (bold line), $\ln
F_{Coulomb}$ (dotted line) and $\ln F_{Ueh}$ (dashed line) versus
$R$ ($0<R<10^{-5}$) for $\hbar =c=1$, $\beta =-1$, $\gamma =-1$,
$r_{c}=-1$, $\alpha =3$, $\alpha ^{\prime }=1/137$ and $m=1$. }
\label{FigS}
\end{figure}
\begin{figure}[tbp]
\epsfxsize=12cm \centerline{\epsffile{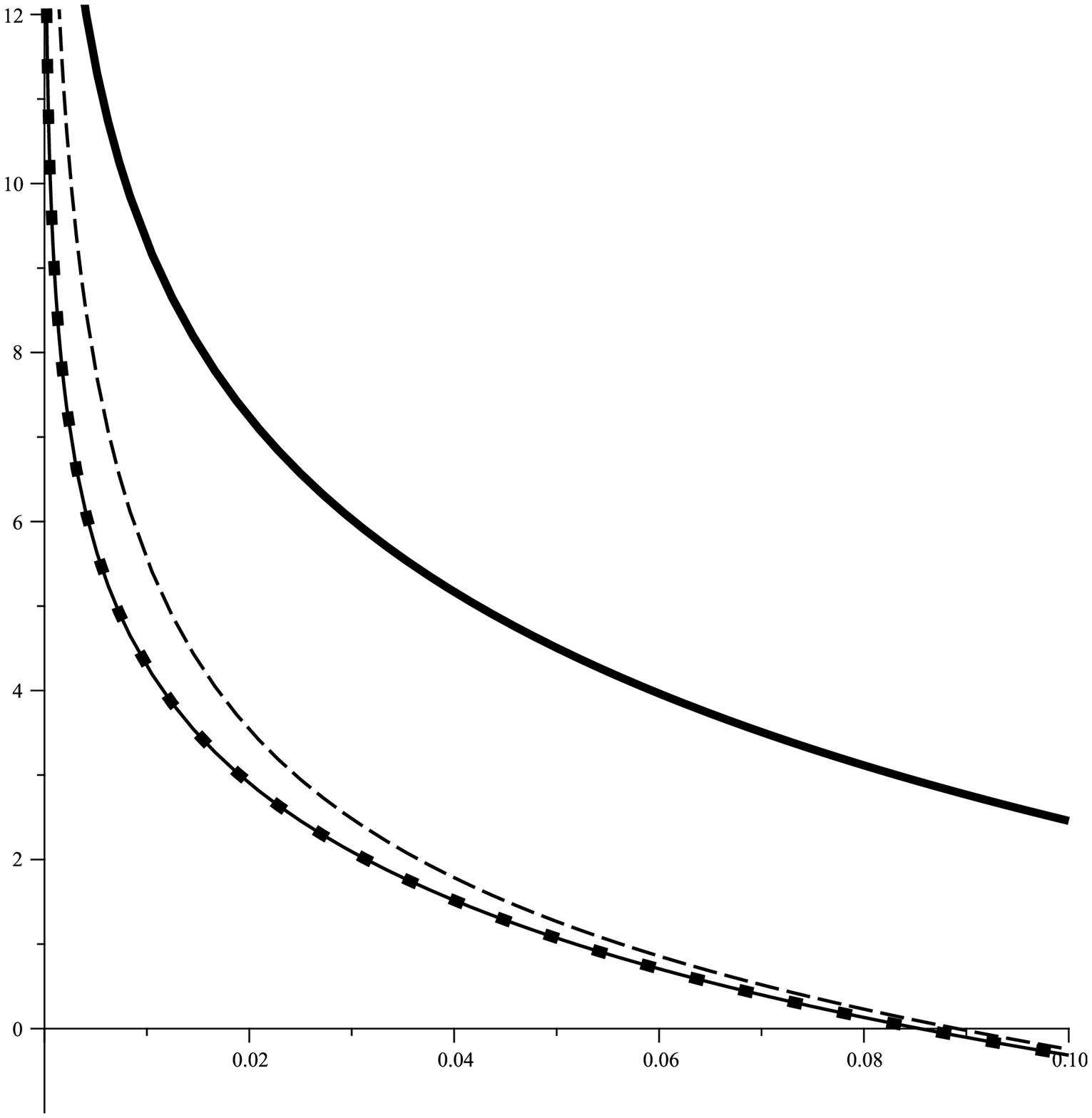}} \caption{$\ln
F_{em1}$ (solid line), $\ln F_{em2}$ (bold line), $\ln
F_{Coulomb}$ (dotted line) and $\ln F_{Ueh}$ (dashed line) versus
$R$\ ($0<R<0.1$) for $\hbar =c=1$, $\beta =-1$, $\gamma =-1$,
$r_{c}=-1$, $\alpha =3$, $\alpha ^{\prime }=1/137$ and $m=1$.}
\label{FigM}
\end{figure}
\begin{figure}[tbp]
\epsfxsize=12cm \centerline{\epsffile{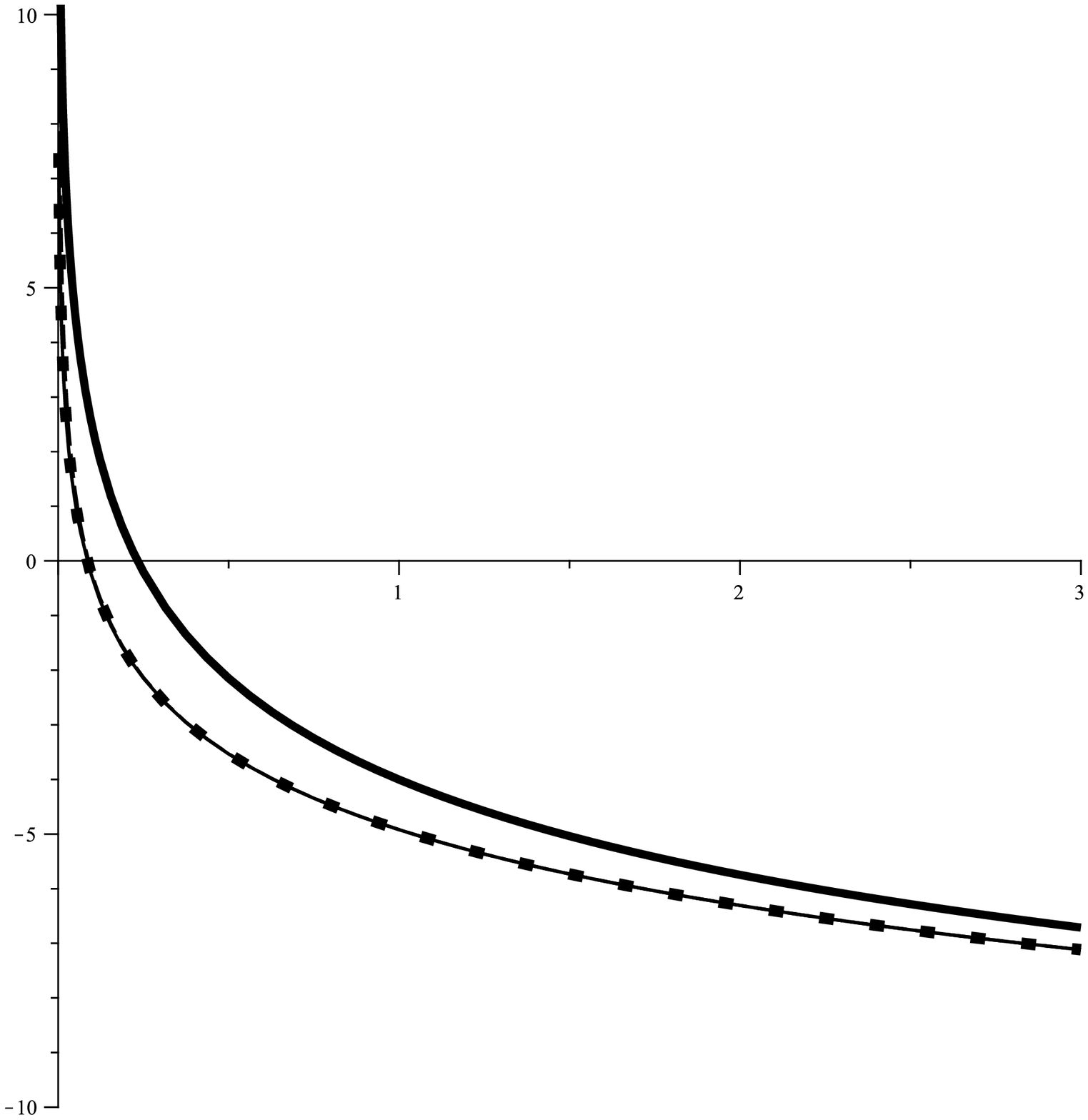}} \caption{$\ln
F_{em1}$ (solid line), $\ln F_{em2}$ (bold line), $\ln
F_{Coulomb}$ (dotted line) and $\ln F_{Ueh}$
(dashed line) versus $R$\ ($0<R<3$) for $\hbar =c=1$, $\beta =-1$, $%
\gamma =-1$, $r_{c}=-1$, $\alpha =3$, $\alpha ^{\prime }=1/137$
and $m=1$.} \label{FigL}
\end{figure}
In order to compare the results of the entropic and the Uehling
corrections, we can plot the corresponding forces for different
values of distance $R$. We draw three logarithmic figures for
different scale. Figure \ref{FigS}, which is drawn for very small
scale ($0<R<10^{-5}$), shows that for small value of $R$,
$F_{em2}$ is compatible with $F_{Ueh}$ and for a special value of
$R$ they are equal, and also $F_{em1}$ is near to the Coloumb
force, $F_{col}$. When we investigate the figure \ref{FigM}, which
is drawn for medium scale ($0<R<0.1$), we find that in this scale,
$F_{em2}$ is far from others. One can find that in figure
\ref{FigL}, which is plotted for large scale ($0<R<3$), $F_{em1}$,
$F_{Ueh}$ and $F_{col}$ are overlapped to each other and $F_{em2}$
is separated. These figures show that for small values of distance
$F_{em2}$ is more compatible with Uehling force, but for large
values of $R$, $F_{em1}$ is more near to $F_{Ueh}$. As a result it
is interesting to study the entropic force arising from the change
of information.

\section{Generalized Equipartition Rule and Newton-Coulomb's Law}

In this section we would like to generalize our discussion in the
previous section to the case where electromagnetic force as well
as the gravitational force are considered. We will study two
approaches in dealing with the problem.

\subsection{First Approach}

In the first approach we identify the total relativistic rest
energy as
\begin{equation}
E=Mc^{2}+\Gamma Qc^{2},  \label{equip0}
\end{equation}%
and thus the equipartition rule (\ref{Ec}) will be replaced with
\begin{equation}
Mc^{2}+\Gamma Qc^{2}=\frac{1}{2}Nk_{B}T,  \label{equip1}
\end{equation}%
Inserting Eqs. (\ref{equip0})- (\ref{equip1}) in Eq. (\ref{F2})
after using Eqs. (\ref{S3}) and (\ref{AQN}), we find
\begin{eqnarray}
F_{\mathrm{g,em}} &=&\frac{2\left( M+\Gamma Q\right) c^{2}}{Nk_{B}}\frac{%
\Delta A}{\Delta x}\left( \frac{\partial S}{\partial A}\right)  \nonumber \\
&=&\frac{2\left( M+\Gamma Q\right) \xi mc^{3}}{Nk_{B}\eta \hbar
}\left(
\frac{\partial S}{\partial A}\right)  \nonumber \\
&=&\frac{\left( mM+\chi qQ\right) }{R^{2}}\left( \frac{\xi
^{2}c^{3}}{8\pi
k_{B}\eta \hbar \ell _{p}^{2}}\right) \left[ 1+4\ell _{p}^{2}\frac{\partial {%
s}}{\partial A}\right] _{A=4\pi R^{2}}  \label{FM&Q}
\end{eqnarray}%
Again if we define $\xi ^{2}=8\pi k_{B}\eta \ell _{p}^{4}$ and
$\chi =k/G=\hbar /(4\pi \varepsilon _{0}\ell _{p}^{2}c^{3})$,
after also using Eqs. (\ref{S1}) and (\ref{plec}), we reach
directly the modified Newton-Coulomb's law corresponding to the
logarithmic and power-law corrections, respectively,
\begin{equation}
F_{\mathrm{g,em}}=\frac{GmM+kqQ}{R^{2}}\left[ 1-\frac{\beta }{\pi }\frac{%
\ell _{p}^{2}}{R^{2}}-\frac{\gamma }{4\pi ^{2}}\frac{\ell _{p}^{4}}{R^{4}}%
\right] ,  \label{FM&Q2}
\end{equation}%
\begin{equation}
F_{\mathrm{g,em}}=\frac{GmM+kqQ}{R^{2}}\left[ 1-\frac{\alpha
}{2}\left( \frac{r_{c}}{R}\right) ^{\alpha -2}\right] .
\label{FM&Q22}
\end{equation}%
These are the total entropy-corrected forces between a test
particle with charge $q$ and mass $m$ in a distance $R$ of a
source particle with charge $Q$ and mass $M$. We see that the
correction terms have the same form for both gravitation and
electromagnetic forces. If one of the particle does not have
charge, i.e. $q=0$ or $Q=0$, then Eq. (\ref{FM&Q2}) reduces to the
quantum correction Newton's law of gravitation \cite{Sheykhi}.
Again we see that the corrections play a significant role only at
the very small distances of $R$.

\subsection{Second Approach}

The second approach is very simple. It is sufficient to add the
modified electromagnetic force obtained in Eq. (\ref{F4}) and the
modified Newton's law of gravitation derived in \cite{Sheykhi},
where in the general form is
\begin{equation}
F_{\mathrm{g}}= G\frac{mM}{R^{2}}\left[ 1+4\ell _{p}^{2}\frac{\partial {s}}{%
\partial A}\right] _{A=4\pi R^{2}}.
\end{equation}%
Since the emergent directions of gravity and electromagnetic
forces coincide, we can obtain
\begin{eqnarray*}
F_{\mathrm{g,em}} &=&F_{\mathrm{g}}+F_{\mathrm{em}} \\
&=& \frac{GmM+kqQ}{R^{2}}\left[ 1+4\ell _{p}^{2}\frac{\partial
{s}}{\partial A}\right] _{A=4\pi R^{2}}.
\end{eqnarray*}%
Using Eqs. (\ref{S1}) and (\ref{plec}), it is straightforward to
recover Eqs. (\ref{FM&Q2}) and (\ref{FM&Q22}).

\section{Entropy corrected Poisson's Equation}

We can also derive the modified Poisson's equation for the
electric
potential $\phi $, provided we define a new wavelength $\lambda _{q}=\frac{%
\delta \hbar }{qc}$ instead of Compton wavelength, $\lambda
_{m}=\frac{\hbar }{mc}$, where $\delta =\sqrt{4\pi \varepsilon
_{0}G}$. This definition may be understood if one accept a
correspondence between the role of mass $m$ in gravitational force
and the role of charge $q$ in the electromagnetic force. Consider
the differential form of Gauss's law
\begin{equation}
\overrightarrow{\nabla }.\overrightarrow{E}=\frac{\rho
}{\varepsilon _{0}},
\end{equation}%
and the fact that electrical field has zero curl and equivalently $%
\overrightarrow{E}=-\overrightarrow{\nabla }\phi $, where $\phi $\
is the electrical potential, it is easy to obtain the familiar
Poisson's equation as
\begin{equation}
\nabla ^{2}\phi =-\frac{\rho }{\varepsilon _{0}}.
\end{equation}%
In this section, by assuming the modified entropy-area relation
(\ref{S2}),
we want to obtain the modified Poisson's equation. It was argued in \cite%
{Ver} that the holographic screens correspond to the equipotential
surfaces, so it
is natural to define%
\begin{equation}
-\frac{\delta N}{2c^{2}}\nabla \phi =\frac{\Delta S}{\Delta x},
\label{P1}
\end{equation}%
where $\frac{\Delta S}{\Delta x}=\left( \frac{\partial S}{\partial
A}\right)
\frac{\Delta A}{\Delta x}$. Substituting $N=\frac{A}{\ell _{p}^{2}}$, $%
\Delta A=\ell _{p}^{2}$ and $\Delta x=\frac{\lambda _{q}}{8\pi }$, where $%
\lambda _{q}=\frac{\delta \hbar }{qc}$ in Eq. (\ref{P1}), we can
rewrite it in the differential from
\begin{equation}
-\frac{\sqrt{\pi \varepsilon _{0}G}}{\ell _{p}^{2}c^{2}}\nabla \phi dA=\frac{%
4\pi c\ell _{p}^{2}}{\sqrt{\pi \varepsilon _{0}G}\hbar }\left( \frac{%
\partial S}{\partial A}\right) dq.  \label{P2}
\end{equation}%
Using the divergence theorem, we find
\begin{equation}
-\frac{\sqrt{\pi \varepsilon _{0}G}}{\ell _{p}^{2}c^{2}}\int
\nabla ^{2}\phi dV=\frac{4\pi c\ell _{p}^{2}q}{\sqrt{\pi
\varepsilon _{0}G}\hbar }\left( \frac{\partial S}{\partial
A}\right) .  \label{P3}
\end{equation}%
Now, we are in a position to extract the modified Poisson's
equation
\begin{equation}
\nabla ^{2}\phi =-\frac{4\pi \ell _{p}^{4}c^{3}}{\pi \varepsilon _{0}G\hbar }%
\left( \frac{\partial S}{\partial A}\right) \frac{dq}{dV},
\label{P4}
\end{equation}%
Using Eq. (\ref{S3}), the above equation can be further rewritten
\begin{equation}
\nabla ^{2}\phi =-\frac{\ell _{p}^{2}c^{3}}{\varepsilon _{0}G\hbar }\rho %
\left[ 1+4l_{p}^{2}\frac{\partial s}{\partial A}\right] ,
\label{ModP0}
\end{equation}
where we have defined the charge density $\rho =dq/dV$. Finally,
using the fact that $c^{3}\ell _{p}^{2}/\hbar =G$, we can write
the modified Poisson's equation in the following manner
\begin{equation}
\nabla ^{2}\phi =-\frac{\rho }{\varepsilon _{0}}\left[ 1+4l_{p}^{2}\frac{%
\partial s}{\partial A}\right] _{A=4\pi R^{2}},  \label{ModP1}
\end{equation}%
where it reduces to
\begin{equation}
\nabla ^{2}\phi =-\frac{\rho }{\varepsilon _{0}}\left[
1-\frac{\beta }{\pi } \frac{\ell _{p}^{2}}{R^{2}}-\frac{\gamma
}{4\pi ^{2}}\frac{\ell _{p}^{4}}{ R^{4}}\right] ,  \label{ModP}
\end{equation}
and
\begin{equation}
\nabla ^{2}\phi =-\frac{\rho }{\varepsilon _{0}}\left[
1-\frac{\alpha }{2} \left( \frac{r_{c}}{R}\right) ^{\alpha
-2}\right] ,  \label{ModP2}
\end{equation}
for logarithmic and power-law corrections, respectively. In this
way, one can derive the quantum correction to Poisson's equation.

\section{Conclusions\label{Sum}}

To conclude, taking into account the quantum corrections in area
law of the black hole entropy, we derived the modified Coulomb's
law of electromagnetic as well as the generalized Newton-Coulomb's
law in the presence of correction terms. In addition we
investigated the vacuum-polarization correction in QED (Uehling
potential) and found that the results of entropic corrections of
Coulomb's law is near to the Uehling potential for some distances.
This compatibility motivated us to investigate the entropic force
in other electromagnetic field equations. We also obtained
entropy-corrected Poisson's equation which governing the evolution
of the scalar potential $\phi$. Our study is the quite one
generalization of Verlinde's argument on the gravity force, to the
electromagnetic interaction. According to the Verlinde's
discussion the gravitational force has a holographic origin. In
this work we proposed a similar nature for the electromagnetic
interaction. Our motivation is the high apparent similarity
between the Newton's law and the Coulomb's law. If gravity and
electromagnetic interaction can be extracted from holographic
principle, this can be regarded as a form unification of gravity
and electromagnetic force. Interestingly enough, we found that the
correction terms have similar form for both Newton's law and
Coulomb's law. This implies that in the very small distances,
these two fundamental forces have the same behavior. This fact
further supports the unification of gravity and electromagnetic
interactions based on the holographic principle.

\section*{Acknowledgements}

We thank the referees for constructive comments. This work has
been supported financially by Research Institute for Astronomy and
Astrophysics of Maragha, Iran.


\begin{thebibliography}{99}
\bibitem{HB} J. D. Bekenstein,  Phys. Rev. D 7,
2333 (1973);

S. W. Hawking, Commun Math. Phys. 43, 199 (1975);

 S. W. Hawking, Nature 248, 30 (1974).

\bibitem{B} J. M. Bardeen, B. Carter and S. W. Hawking, Commun. Math. Phys. 31,
161 (1973).

\bibitem{D} P. C. W. Davies, J. Phys. A: Math. Gen. 8, 609 (1975);

W. G. Unruh, Phys. Rev. D 14, 870 (1976);

L. Susskind, J. Math. Phys. 36, 6377 (1995).

\bibitem{Jac} T. Jacobson, Phys. Rev. Lett. \textbf{75}, 1260 (1995).

\bibitem{Pad1} T. Padmanabhan, Class. Quantum. Grav. \textbf{19}, 5387
(2002).

\bibitem{Elin} C. Eling, R. Guedens and T. Jacobson, Phys. Rev. Lett.
\textbf{96}, 121301 (2006).

\bibitem{Cai2} M. Akbar and R. G. Cai, Phys. Rev. D \textbf{75}, 084003
(2007).

\bibitem{Cai3} R. G. Cai and L. M. Cao, Phys.Rev. D \textbf{75}, 064008
(2007).

\bibitem{CaiKim} R. G. Cai and S. P. Kim, JHEP \textbf{0502}, 050 (2005).

\bibitem{Wang} B. Wang, E. Abdalla and R. K. Su, Phys. Lett. B \textbf{503},
394 (2001);

B. Wang, E. Abdalla and R. K. Su, Mod. Phys. Lett. A \textbf{17},
23 (2002).

\bibitem{Cai33} R. G. Cai, L. M. Cao and Y. P. Hu, JHEP \textbf{0808}, 090
(2008).

\bibitem{Shey0} S. Nojiri and S. D. Odintsov, Gen. Relativ. Gravit. \textbf{%
38}, 1285 (2006);

A. Sheykhi, Class. Quantum Grav. \textbf{27}, 025007 (2010);

A. Sheykhi, Eur. Phys. J. C \textbf{69}, 265 (2010).

\bibitem{Shey1} A. Sheykhi, B. Wang and R. G. Cai, Nucl. Phys. B \textbf{779}, 1 (2007);

R. G. Cai and L. M. Cao, Nucl. Phys. B \textbf{785}, 135 (2007).

\bibitem{Shey2} A. Sheykhi, B. Wang and R. G. Cai, Phys. Rev. D \textbf{76},
023515 (2007);

A. Sheykhi, B. Wang, Phys. Lett. B \textbf{678}, 434 (2009).

\bibitem{Padrev} T. Padmanabhan, Rept. Prog. Phys. \textbf{73}, 046901
(2010).

\bibitem{Ver} E. P. Verlinde, JHEP \textbf{1104}, 029 (2011).


\bibitem{Hossenfelder2010} S. Hossenfelder, [arXiv:1003.1015];

A. Kobakhidze, Phys. Rev. D 83, 021502 (2011);

B. L. Hu, Int. J. Mod. Phys. D 20, 697 (2011);

A. Kobakhidze, [arXiv:1108.4161].

\bibitem{Gao2011} S. Gao, Entropy, 13, 936 (2011).

\bibitem{Winkelnkemper} H. E. Winkelnkemper, AP Theory V: Thermodynamics in Topological Disguise, Gravity from
Holography and Entropic Force as Dynamic Dark Energy. Available
online:http://www.math.umd.edu/~hew/ (accessed on 7 April 2011);
Preprint, February 2011

\bibitem{Myung2011} Y. S. Myung, Eur. Phys. J. C 71, 1549
(2011).

\bibitem{Chaichian2011} M. Chaichian, M. Oksanen and A. Tureanu, [arXiv:1109.2794].

\bibitem{Cai4} R. G. Cai, L. M. Cao and N. Ohta, Phys. Rev. D \textbf{81},
061501(R) (2010);

Y. Ling and J. P. Wu, JCAP \textbf{1008}, 017 (2010).

\bibitem{Sheykhi} A. Sheykhi, Phys. Rev. D \textbf{81}, 104011 (2010);

A. Sheykhi and S. H. Hendi, Phys. Rev. D \textbf{84}, 044023
(2011).

\bibitem{Modesto} L. Modesto and A. Randono, [arXiv:1003.1998].

\bibitem{smolin} L. Smolin, [arXiv:1001.3668].

\bibitem{Mli} M. Li and Y. Wang, Phys. Lett. B \textbf{687}, 243
(2010);

D. A. Easson, P. H. Frampton and G. F. Smoot, Phys. Lett. B
\textbf{696}, 273 (2011);

U. H. Danielsson, [arXiv:1003.0668].

\bibitem{Tian} Y. Tian and X. Wu, Phys. Rev. D \textbf{81}, 104013 (2010);

\bibitem{Twang} T. Wang, Phys. Rev. D \textbf{81}, 104045 (2010).

\bibitem{Other} Y. X. Liu, Y. Q. Wang, S. W. Wei, Class. Quantum Grav. \textbf{27}, 185002 (2010);

V. V. Kiselev and S. A. Timofeev, Mod. Phys. Lett. A \textbf{25},
2223 (2010);

R. A. Konoplya, Eur. Phys. J. C \textbf{69}, 555 (2010);

R. Banerjee and B. R. Majhi. Phys. Rev. D \textbf{81}, 124006
(2010);

P. Nicolini, Phys. Rev. D \textbf{82}, 044030 (2010);

C. Gao, Phys. Rev. D \textbf{81}, 087306 (2010);

Y. S. Myung and Y.W Kim, Phys. Rev. D \textbf{81}, 105012
(2010);

H. Wei, Phys. Lett. B \textbf{692}, 167 (2010);

D. A. Easson, P. H. Frampton and G. F. Smoot,
[arXiv:1003.1528];

S. W. Wei, Y. X. Liu and Y. Q. Wang, \emph{to be published in
Commun. Theor. Phys.} [arXiv:1001.5238].

\bibitem{Meis} K. A. Meissner, Class. Quantum Grav. \textbf{21}, 5245
(2004);

A. Ghosh and P. Mitra, Phys. Rev. D \textbf{71}, 027502 (2004);

A. Chatterjee and P. Majumdar, Phys. Rev. Lett. \textbf{92},
141301 (2004).

\bibitem{Zhang} J. Zhang, Phys. Lett. B \textbf{668}, 353 (2008);

R. Banerjee and B. R. Majhi, Phys. Lett. B \textbf{662}, 62
(2008);

R. Banerjee and B. R. Majhi, JHEP \textbf{0806}, 095
(2008);

S. Nojiri and S. D. Odintsov, Int. J. Mod. Phys. A \textbf{16},
3273 (2001).

\bibitem{Sau} S. Das, S. Shankaranarayanan and S. Sur, Phys. Rev. D \textbf{%
77}, 064013 (2008).

\bibitem{sau1} S. Das, S. Shankaranarayanan and S. Sur, [arXiv:1002.1129];

S. Das, S. Shankaranarayanan and S. Sur, [arXiv:0806.0402].

\bibitem{pavon1} N. Radicella, D. Pavon, Phys. Lett. B \textbf{691}, 121
(2010).

\bibitem{Pad3} T. Padmanabhan, Class. Quantum Grav. \textbf{21}, 4485 (2004);

T. Padmanabhan, Mod. Phys. Lett. A \textbf{25}, 1129 (2010);

T. Padmanabhan, Phys. Rev. D \textbf{81}, 124040 (2010).

\bibitem{Uehling} E. A. Uehling, Phys. Rev. \textbf{48}, 55 (1935);

E. H. Wichmann and N. H. Kroll, Phys. Rev. \textbf{101}, 843
(1956);

A. Bonanno and M. Reuter Phys. Rev. D \textbf{62}, 043008 (2000);

W. Dittrich and M. Reuter, \textit{Effective Lagrangians in
Quantum Electrodynamics}, Springer-Verlag (1985).

\bibitem{Peskin} M. E. Peskin and D. V. Schroeder, \textit{An Introduction
to Quantum Field Theory}, Reading, USA: Addison-Wesley (1995).

\end{thebibliography}
\end{document}